\documentclass[12pt]{article}
\usepackage[dvips]{graphicx}
\usepackage[usenames]{color}
\usepackage{latexsym}
\begin{document}
\title{
Self-organizing social hierarchy and villages
in a challenging society
}
\author{Masaru Tsujiguchi and Takashi Odagaki$^*$\\
Department of Physics, Kyushu University, Fukuoka 812-8581, Japan}
\date{ }
\maketitle
\medskip
\begin{abstract}
We show by Monte Calro (MC) simulation that the hierarchy and villages
emerge simultaneously in a challenging society when the
population density exceeds a critical value.
Our results indicate that among controlling processes of
diffusion and fighting of individuals and relaxation of
wealth, the trend of individuals challeninging to stronger neighbors plays
the pivotal role in the self-organization of the hierarchy and villages.
\end{abstract}


\section{Introduction}

Social structure in various forms exists in the human society and in animals. 
In the Middle Ages, many villages existed each of which was ruled by a feudal
lord and his clan. At present, several nations dominate the world with many
followers and some challengers. A key question is how to understand the
universal nature in the emergence of these hierarchies which consist of a small
number of winners and many losers. It is also an important question to find
the mechanism for the simultaneous emergence of the villages and the hierarchy.
 
Basically, social difference occurs when two moving individuals meet and fight
each other where the winner deprives the loser of wealth or power.
The winning probability of a fight depends on the difference between wealth of
two individuals engaging in the fight. Furthermore, the wealth of an
individual decays to and the negative wealth (debt) increases to zero when
the individual does not fight. Many aspects of the society can be modeled by
setting rules to diffusion, fighting and relaxation processes.

In this paper, we consider a challenging,or bellicose society where individuals try to
challenge thier neibours if possible. We show by Monte Carlo (MC) simulation
that the critical population density for emergence of the hierarchy
is much lower than those in the
no-preference society\cite{bonabeau} and in a timid society\cite{oda-tsuji}.
Furthermore, we show that the hierarchy and villages emerge simultaneously
in this society; in the no-preference society\cite{bonabeau}
or in a timid society\cite{oda-tsuji},
the hierarchy emerges spontaneously but no villages are observed. Namely,
we show that among controlling processes, the trend of individuals
challenging to stronger neighbors plays the critical role
in the self-organization of the structure.

We organize this paper as follows; in Sec. 2, a challenging society is
modelled by setting hostile move of individuals.
The results of the MC simulation is presented in Sec. 3 where the density
dependence of the order parameter and the profile of winning probability.
We also show the formation of villages in the challenging society.
Section 4 is devoted to discussion.

\section{A challenging society}
Bonabeau {\it et al.}\cite{bonabeau}
have shown that a hierarchical society can emerge spontaneously from an
equal society by a simple algorithm of fighting between individuals
who diffuse on a square lattice by a one step simple random walk.
Suppose individual $i$ tries to move onto the site occupied by individual $j$
and these two individuals engage in a fighting.
The fighting rule is characterised by the winning probability $w_{ij}$ of
individual $i$ against individual $j$ which is assumed to be
\begin{equation}
w_{ij} = \frac{1}{1+\exp\{\eta(F_j-F_i)\}} ,
\end{equation}
where $F_i$ is the
wealth of individual $i$ and $\eta (>0)$ is a controlling parameter of
the model.
Therefore, when the difference of the wealths is large, the stronger one wins
all the fights, and when $F_i \simeq F_j$, the winning probability deviates
from $1/2$ linearly in the difference $F_i-F_j$. The winner occupies the
lattice site and increases its wealth by 1, and the loser moves to the site
previously occupied by $i$ and reduces its wealth by 1.

When individual $i$ is not involved in any fight in one MC time step
(MC tries during which all idividuals are accessed once),
its wealth is assumed to decay as
\begin{equation}
F_i(t+1) = F_i(t) - \mu \tanh [F_i(t)] ,
\end{equation}
where the unit of time is one MC step.
When the wealth is large, it decays by a constant amount per one MC step,
$F_i(t+1) = F_i(t) - \mu$, i.e. a rich person does not waste his/her
wealth. When the wealth is small, it decreases at a constant rate,
that is $F_i(t+1) = (1 - \mu)F_i(t)$.  Here, $\mu$ is another controlling
parameter of the model.

The social hierarchy can be characterized by the fact that some people have
won and some other people have lost more fights. Suppose individual $i$
won $W_i$ times in $X_i$ fights for a given time interval.
Then the order parameter $\sigma$ can be defined by the mean square deviation
of $W_i/X_i$ from $1/2$, 
\begin{equation}
\sigma^2 = \frac{1}{N}\sum_i\biggl\{\frac{W_i}{X_i}-\frac{1}{2}\biggr\}^2 .
\end{equation}
Bonabeau {\it et al} showed by MC
simulation that the social hierarchy self-organizes at a critical density
as the population density is increased. Note that the relaxation process plays
a critical role to have such a transition\cite{sousa, stauffer}.

In order to study the emergence of social hierarchy and villages in the
society of challengers, we introduce a bellicose diffusion strategy:
When an individual makes one step random walk on the square lattice,
it always moves to a site occupied by some one, and when more than two sites
are occupied, it always challenges the strongest among them.
An individual is prohibited to fight suscessively with the same opponent.
Employing the same rule for the fighting and relaxation processes as
Bonabeau {\it et al}\cite{bonabeau}, we examined the emergence of hierarchy
and spacial structure in this society by MC simulation.

\section{Monte Carlo simulation}
MC simulation was performed for $N = 3500$ individuals on the $L \times L$ square lattice
with periodic boundary conditions from $L = 60$ to $L = 600$.

Figure 1 shows the dependence of the order parameter on the population density.
We see the transition occurs at $\rho \simeq 0.04$ when $\mu=0.1$ and $\eta = 0.05$,
which is much lower than the critical value for no-preference society
($\rho \simeq 0.1$ for the same $\mu$ and $\eta$) studied by
Bonabeau {\it et al}\cite{bonabeau}.

\begin{figure}[thb] 
\begin{center} 
\includegraphics[height=4cm]{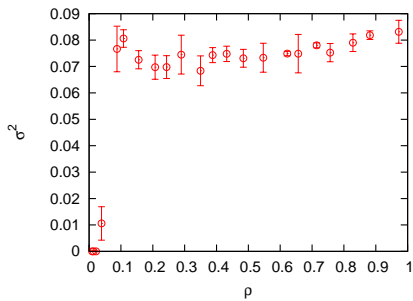} 
\end{center} 
\caption{Order parameter $\sigma ^2$ as a function of $\rho=N/L^2$ for
 $\mu=0.1$ and $\eta = 0.05$.
}
\end{figure} 

The detailed structure in population is monitored by the profile of
the winning frequency $W_i/X_i$.
Figure 2 shows the profile of the winning frequency for four different
population density; $\rho = 0.022$, $0.056$, $0.086$ and $0.714$.
In the egalitarian society at low densities below the critical density,
the profile shows a sharp peak at $W_i/X_i= 0.5$.
When the density exceeds the critical value, the distribution of the
winning probability becomes widespread,  and at the same time
individuals with winning probability above 95\% and with winning
probability less than 5\% emerge,
\begin{figure}[bht] 
\begin{center} 
\includegraphics[height=4cm,width=5cm]{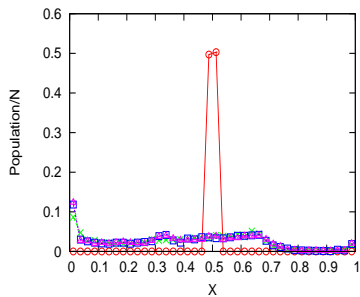} 
\end{center} 
\caption{The profile of the winning frequency for 
four different densities $\rho = 0.022 (\bigcirc)$, $0.056 (\times)$,
$0.086 (\Box)$ and $0.714 (\triangle)$. ($\mu=0.1$ and $\eta = 0.05$.)}
\end{figure} 

\begin{figure}[hbt] 
\begin{center} 
\includegraphics[height=4cm,width=5cm]{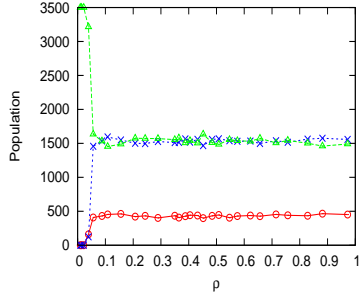} 
\end{center} 
\caption{Dependence of the population in each class on the density
when $\mu = 0.1 $ and $\eta = 0.05$. Winners $(\bigcirc)$, 
losers$(\times)$ and middle class $(\triangle)$.
}
\end{figure} 

We conventionally classify individuals into three groups by the number of
fights which an individual won; winners are individuals who won more
than 2/3 of fights and losers are individuals who won less than 1/3 of fights.
Individuals between these two groups are called middle class.
Figure 3 shows the population of each class as a function
of the population density. It is interesting to note that the emergence of the
hierarchy is signified by appearance of small number of winners.
This is a clear contrast to a timid society where individuals always
avoid fighting\cite{oda-tsuji}. In the timid society, the hierarchical society
emerges in two steps; the first and the second transition are signified by
appearance of losers and winners, respectively.

\begin{figure}[bth] 
\begin{center} 
\includegraphics[height=3cm,width=5cm]{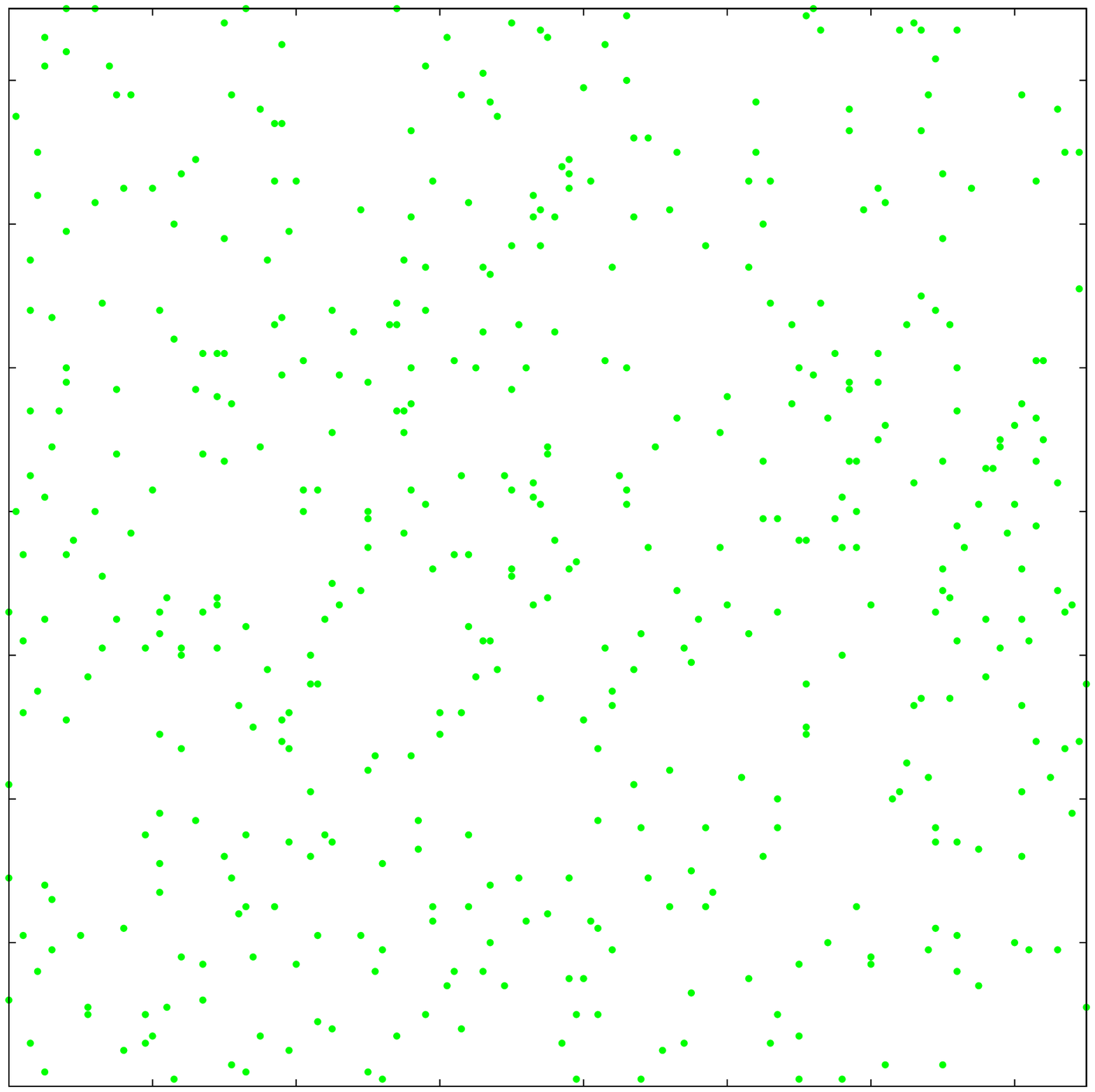} \hspace{0.5cm}
\includegraphics[height=3cm,width=5cm]{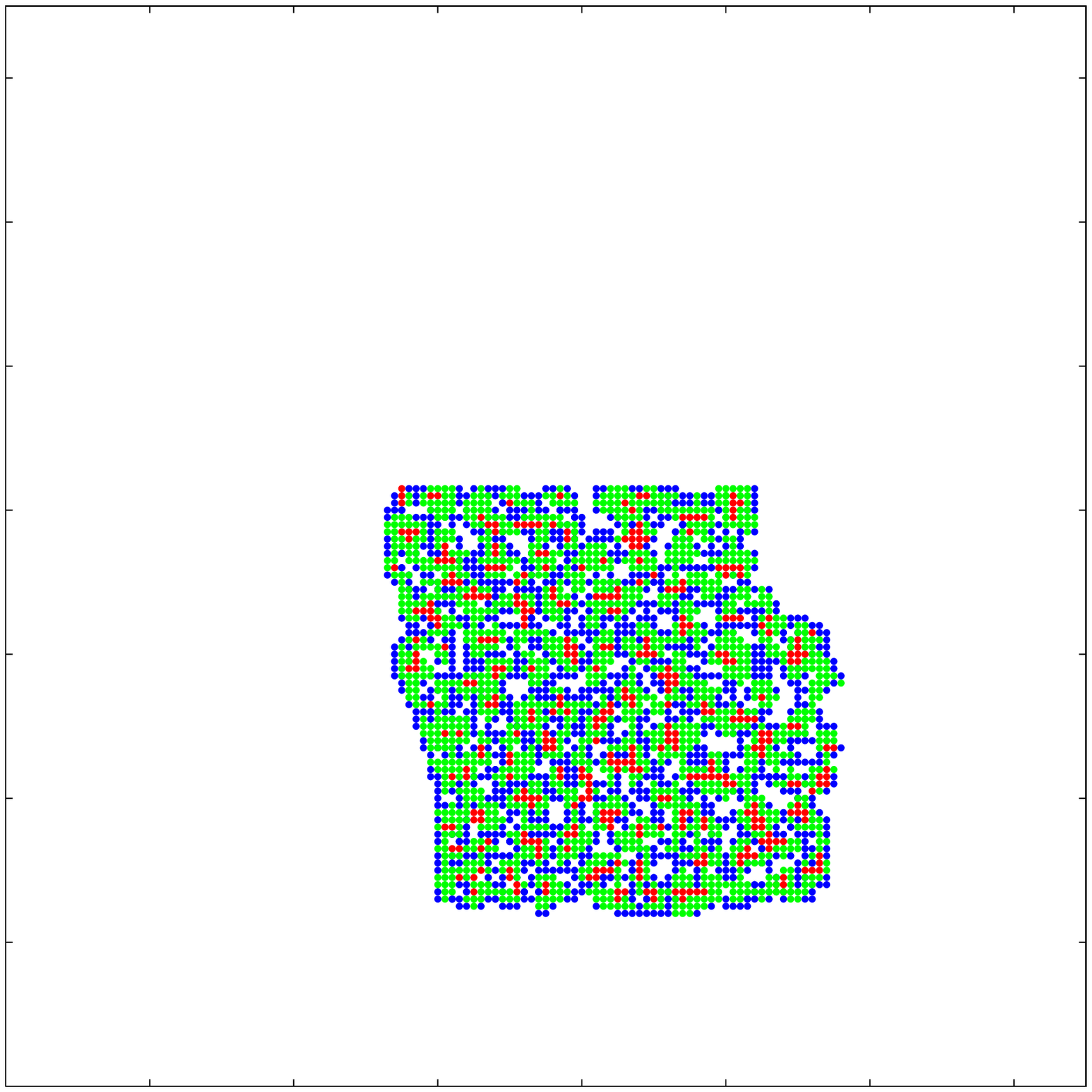} 

\hspace*{-1cm}(a) \hspace{5cm}(b)
\vspace{0.3cm}

\includegraphics[height=3cm,width=5cm]{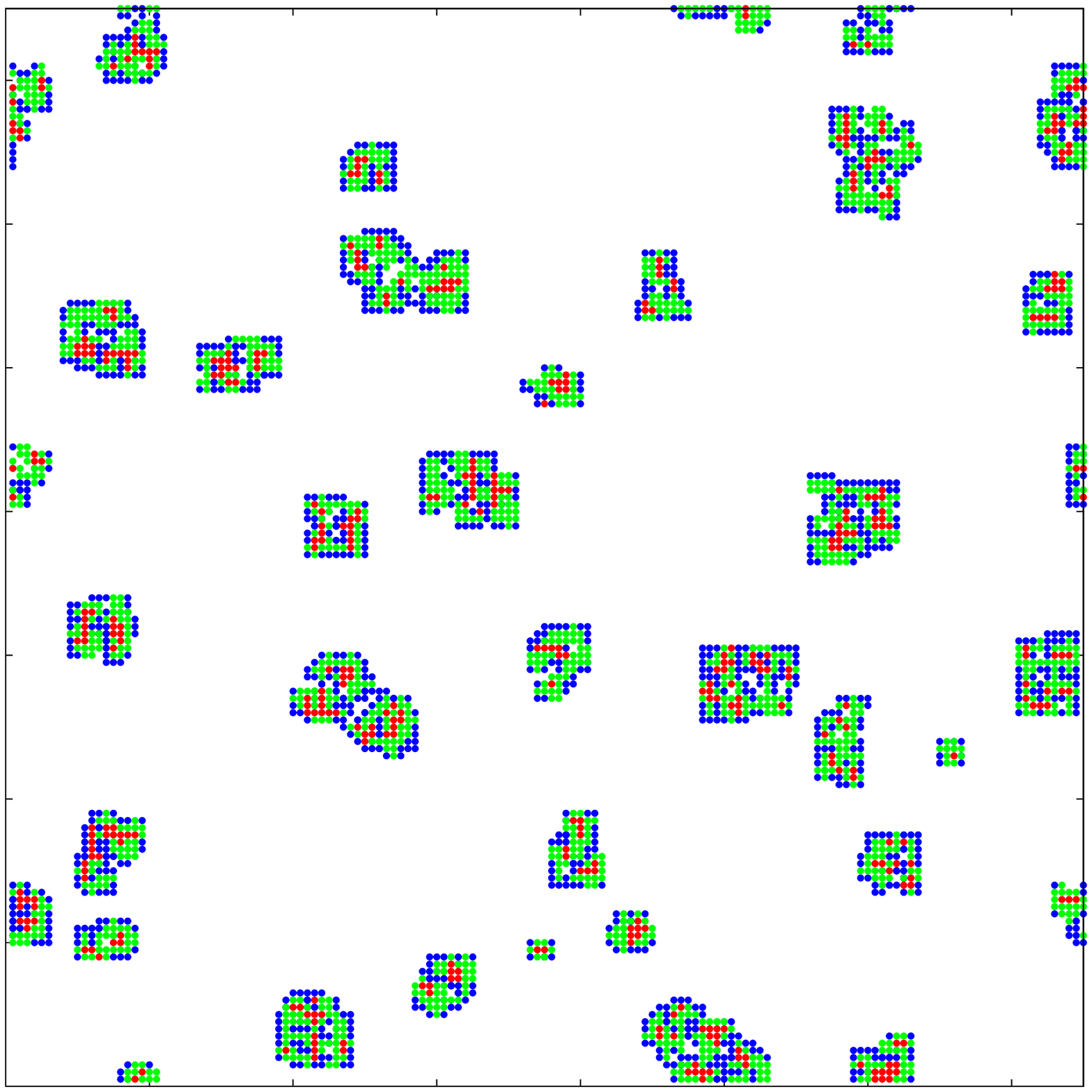}  \hspace{0.5cm}
\includegraphics[height=3cm,width=5cm]{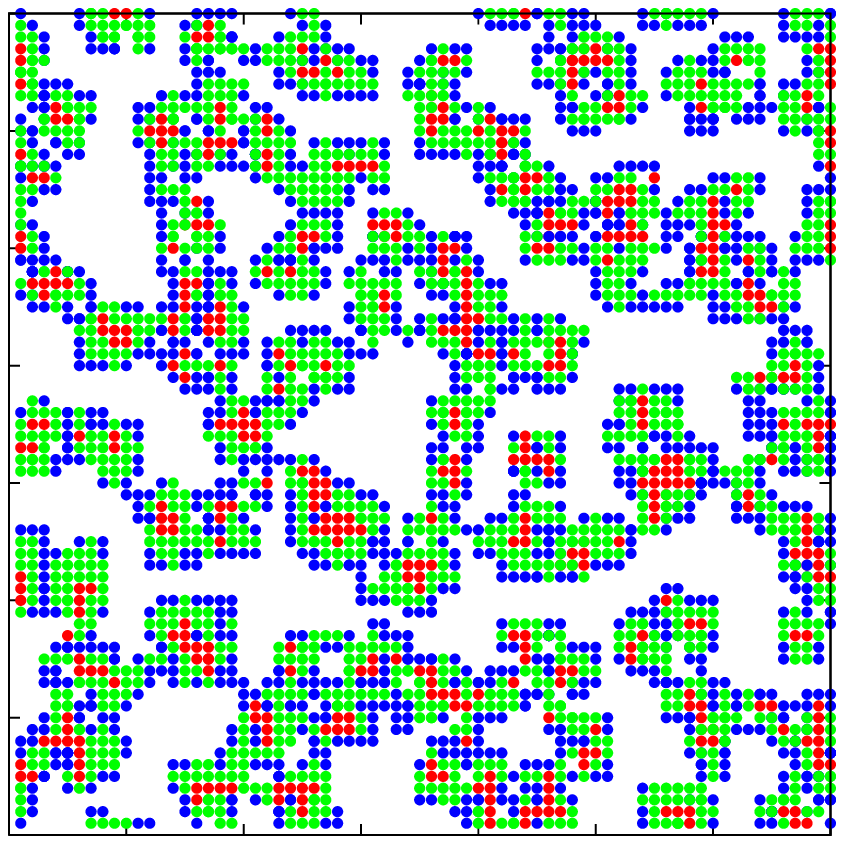}
 
\hspace*{-1cm}(c) \hspace{5cm}(d)
\end{center} 
\caption{Self-organization of villages in the challenging
society when  $\mu = 0.1 $ and $\eta = 0.05$. 
(a) No villages appear at $\rho=0.022$., (b) One big village
is formed at $\rho = 0.056$. (c) Many villages appear at $\rho = 0.086$.
(d) Villages form a percolating cluster at $\rho = 0.714$.
 Winners, losers and middle class are represented by red, blue and green dots, 
respectively.}
\end{figure} 

We now proceed to examine the spatial structure of each state in
the steady state, which is shown in Fig. 4. 
In the egalitarian society, no spatial structure is observed.
When the population density exceeds the critical value,
villages emerge, each of which consists of small number of winners and
large number of middle class and losers. The size of the largest village
depends strongly on the density; At the density just above the critical value,
all individuals belong to one compact cluster as shown in Fig. 4(b).
As the density is increased, the number of clusters increases and thus
the size of the largest cluster is rather small (Fig. 4 (c)).
When the density is larger than a critical percolation density,
one large cluster appears which percolates the system (Fig. 4(d)).
The critical percolation density is about 0.65, which is larger than
the critical percolation density 0.593 of the square lattice.
This is due to the fact that in the model under consideration individuals
have effectively strong attractive interaction\cite{duckers}.

We see that winners (red dots) are near the center of the village,
surrounded by people in the middle class (green dots), and losers
(blue dots) are at its perimeter.
For $\rho = 0.086$, we compare the population profile of winning frequency
of each village, which is shown in Fig. 5
\begin{figure}[htb] 
\begin{center} 
\includegraphics[height=4cm,width=5cm]{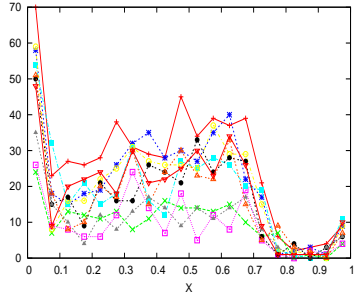} 
\end{center} 
\caption{Population profile of winning frequency of
each village at $\rho =0,086$.  $\mu = 0.1 $ and $\eta = 0.05$.
}
\end{figure} 
It is interesting to observe that the profile is more or less common for
all villages. This may be compared with the structure of medieval villages,
where a few people dominate the village with many subordinates.
The number of villages observed in the observation time
depends on the population density. At higher densites,
villages form a percolating cluster, corresponding to the borderless situation.

\section{Discussion}
We have shown that in a bellicose society the hierarchy self-organizes
at much lower population density compared with the no-preference or a pacifist
societies. Among the basic processes of diffusion, fighting and relaxation,
a small change in the diffusion process affects significantly the
self-organiztion of the social structure. In particluar, preference in the
diffusion process plays an important role in the formation of spatial
structure.
The reason for the villages to be formed in the bellicose society
is in the effective attraction between individuals due to the diffusion
algorithm, namely an individual always stay in the visinity of other
individulas.
Therefore the formation of villages is somewhat similar to the condensation
of droplets in a gas.

In this paper, we have discussed the emergence of villages in the time
period of our MC simulation. It is an open question to find out the
distribution of villages in the long time limit. In fact, there are no
mechnism to keep the center of mass of each village at the same position and
thus each village can diffuse and may collide and merge with other village.

Another open and important quesition is to see the effect of the range of the
random walk. The distance of one step of the random walk represents the mode of
transportation. Therefore, as the mode of transportation advances, 
the effective population density is considered to increase and thus the
globalization may occur at lower population density.
These questions will be studied in the future.
One can expect that various structures of
society can be analyzed within the same frame work,  which will
eventually help in proposing the right policy.




\end{document}